# The Design and Implementation of a Real Time Visual Search System on JD E-commerce Platform

Jie Li, Haifeng Liu, Chuanghua Gui, Jianyu Chen, Zhenyun Ni, Ning Wang
JD.com Inc.
Beijing, China

Yuan Chen
JD.com Silicon Valley R&D Center
Mountain View, CA, USA

## ABSTRACT

We present the design and implementation of a visual search system for real time image retrieval on JD.com, the world's third largest and China's largest e-commerce site. We demonstrate that our system can support real time visual search with hundreds of billions of product images at sub-second timescales and handle frequent image updates through distributed hierarchical architecture and efficient indexing methods. We hope that sharing our practice with our real production system will inspire the middleware community's interest and appreciation for building practical large scale systems for emerging applications, such as e-commerce visual search.

## 1 Introduction

Visual search or image content based retrieval [1-8, 24-30] is a very active area driven by the rapid progress of deep natural network technology for image recognition [5-7]. It is particularly useful for both online and in-store shopping for it can greatly improve customers' experience and engagement. Although significant progress has been made, such as Google Similar Images and Amazon Flows for Internet search, and Pinterest Lens and Google Lens for similar product search, building end-to-end Internet-scale visual search systems for retailing services is still a very challenging task for the following reasons.

First, the product image collection on e-commerce sites is rapidly and continuously growing. There are more than 100 billion product images at JD.com. The visual search system must be able to handle a very large volume of images in a scalable way.

Second, as products are updated very frequently on e-commerce sites, the associated image collection needs to be updated accordingly in a timely and effective matter. At JD, there are about one billion image updates every day. Data freshness becomes an important requirement. The search results should reflect the most recent updates in products or images. Unlike typical internet or web visual search, which builds and updates indexes in minutes, hours or days, and for which relatively outdated information is acceptable, e-commerce visual search has a much more stringent requirement on real time and consistency. This is a critical and unique challenge for our work.

In this paper, we present a scalable real time visual search system on JD.com's e-commerce platform. We make three main contributions.

1. We design and implement a distributed hierarchical system that provides scalable image feature extraction, indexing and retrieval for very large scale image-content based search in real time.
2. We propose a high performance real time image indexing method that utilizes various optimization techniques to support sub-second image update and retrieval.
3. We evaluate the performance of the proposed visual search system on JD.com.

The rest of the paper is organized as follows. Section 2 describes the design and implementation of the proposed visual search system. Section 3 presents the operation and performance data and search examples from the production system. Section 4 reviews the related work and finally Section 5 concludes the paper.

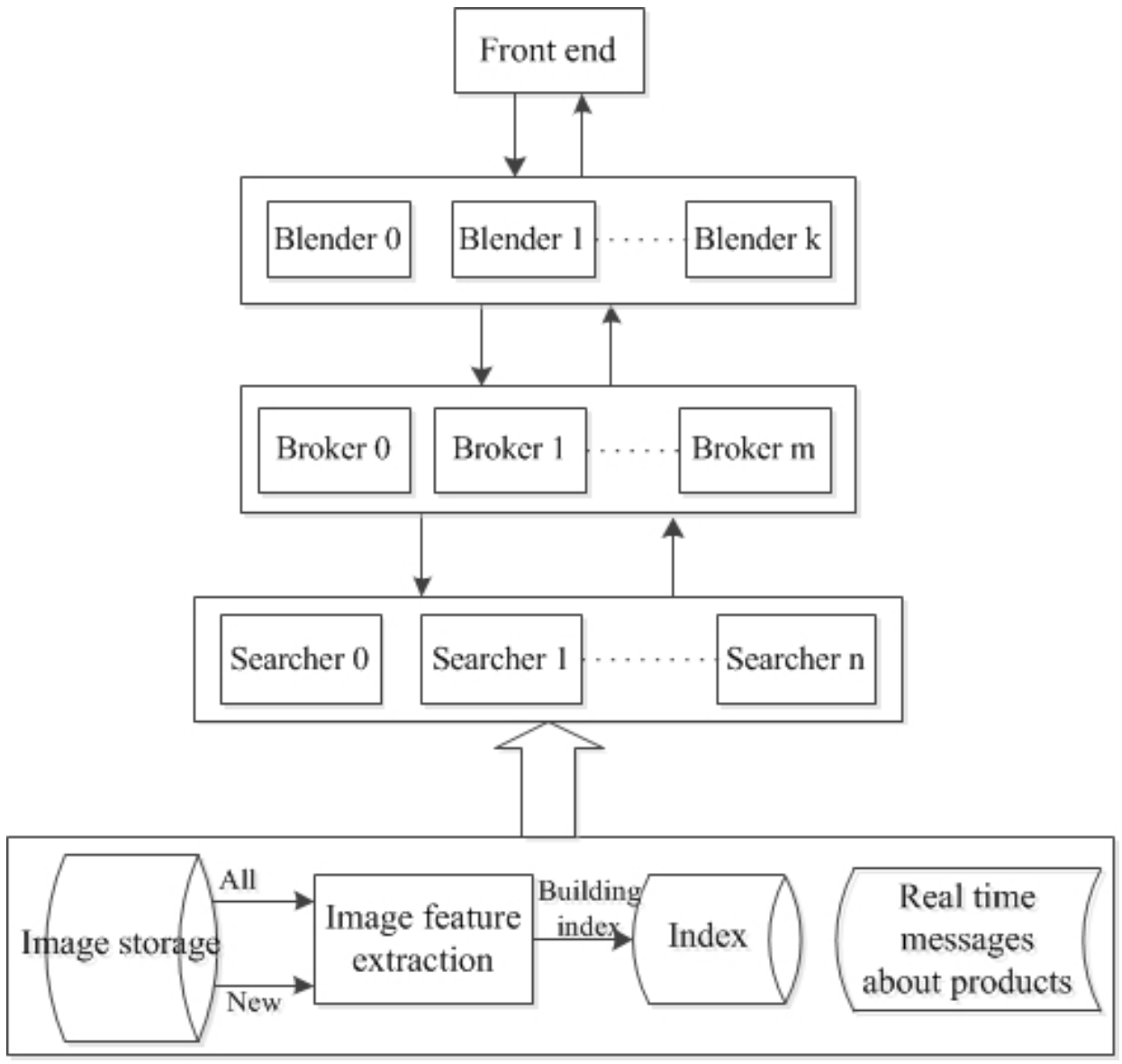

**Figure 1. System Architecture**

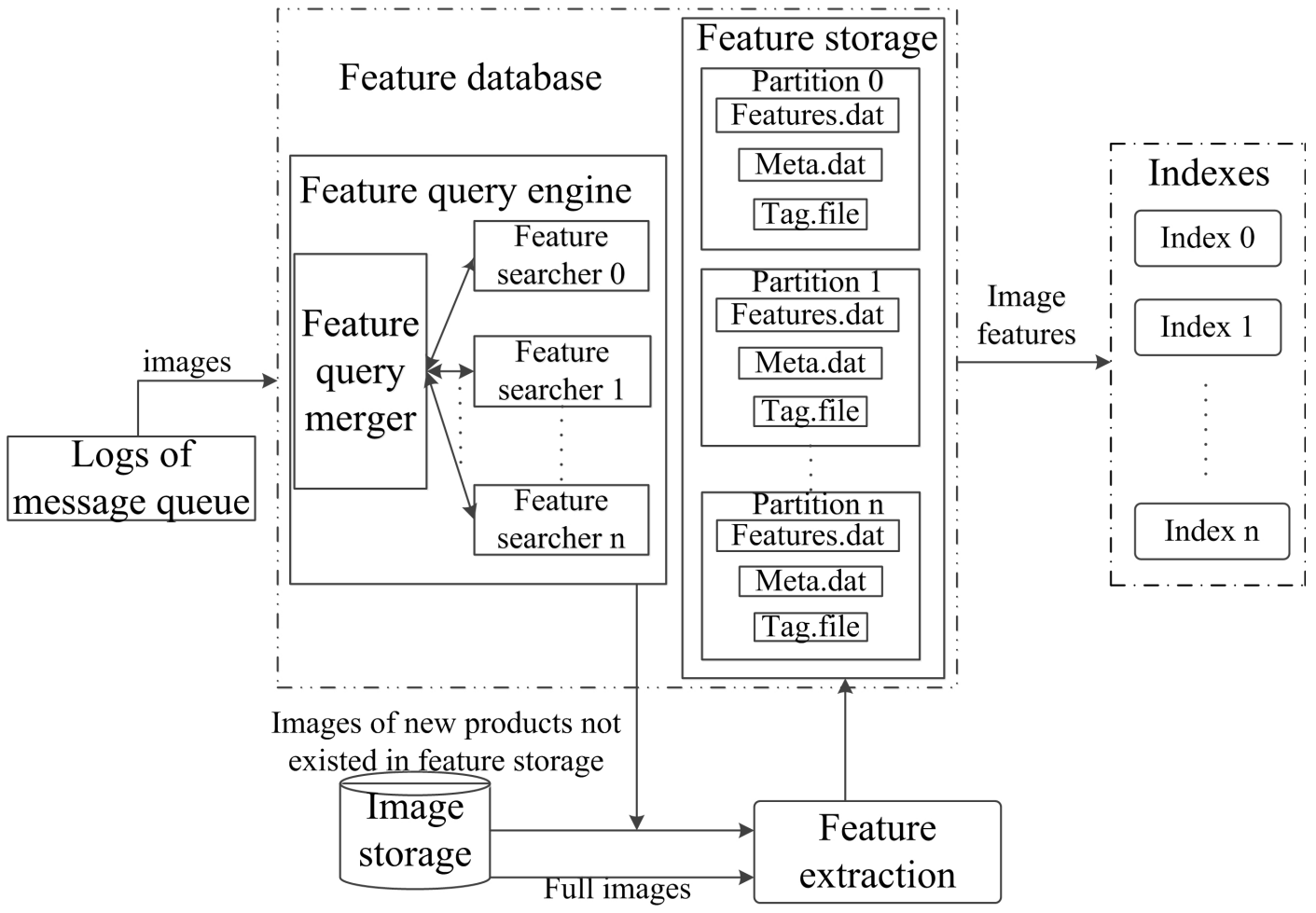

**Figure 2. Full Indexing**

## 2 Design and Implementation

### 2.1 Overview

The architecture of the proposed visual search system is shown in Figure 1. It consists of two subsystems: indexing and search.

**Indexing.** The indexing sub-system is the core of our visual search system. It extracts image features and builds indexes for image search. To ensure data completeness and freshness, both periodic full indexing and real time incremental indexing are performed. The full indexing periodically builds indexes for all the images. The real time indexing is immediately triggered upon receiving a product update event such as addition, deletion and modification of a product. Real time indexing is very critical for good user experience by making sure the search results reflect the most recent product information at the time of reporting. Our real time indexing introduces efficient index structures and optimization methods for image lookup, update, insertion and deletion to improve the performance. The design and implementation detail is discussed in Section 2.2 and 2.3.

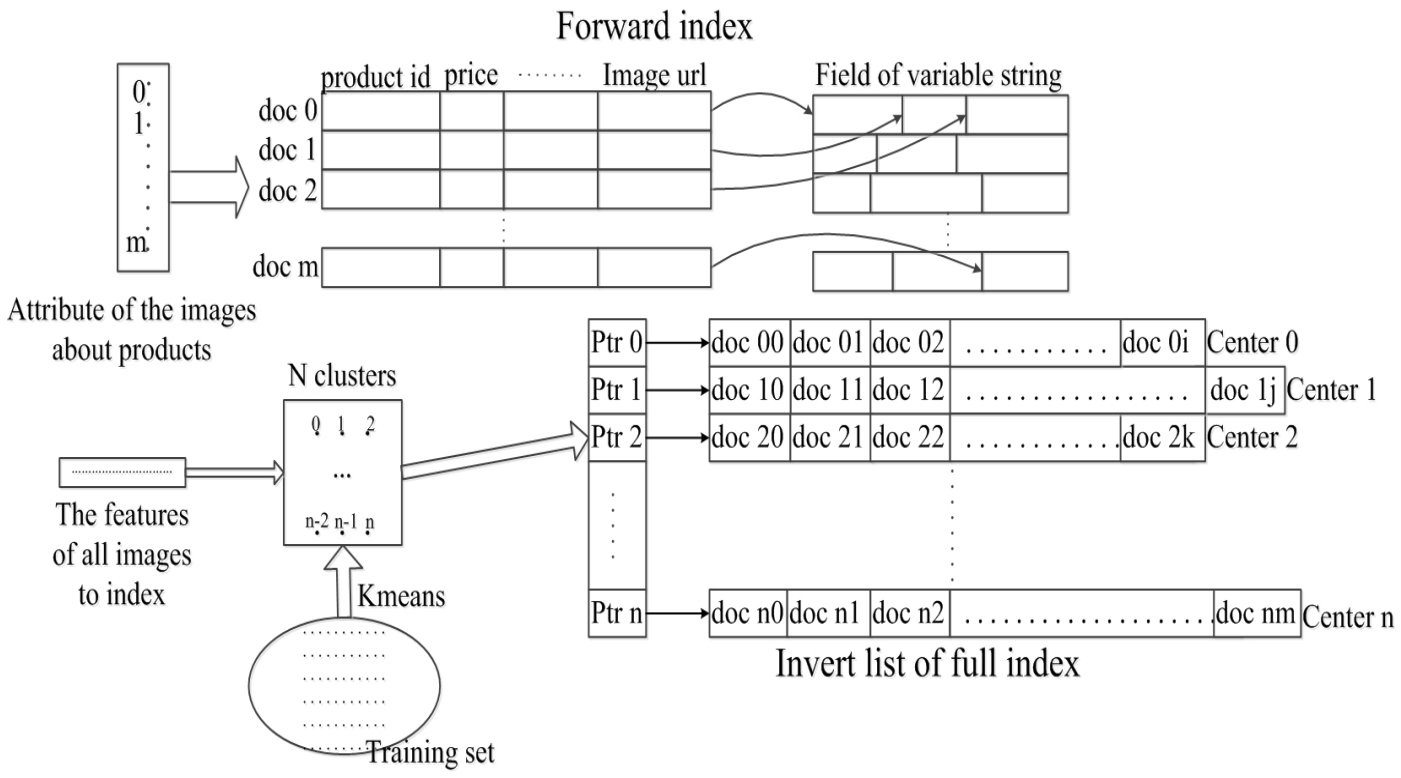

**Figure 3. Building Full Index**

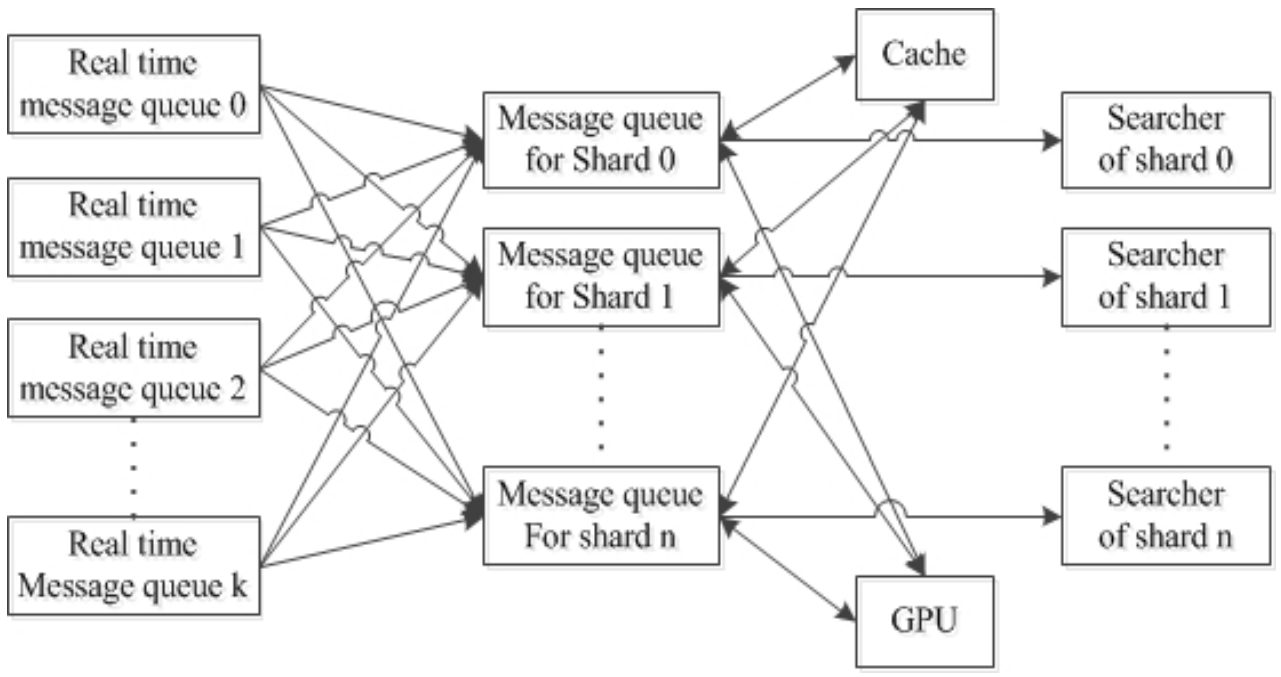

**Figure 4. Real Time Indexing**

E-commerce sites often remove a product from the market and put it back later. The reuse of the product and image's information is critical for good performance, especially for image feature exaction, which is an expensive operation. Our system always checks if an image's features have been previously extracted to avoid the repeated feature extraction. A bitmap is used to indicate if a product or image is valid or not. When a product is removed from the market (e.g., out of stock or other reasons), it is marked invalid and excluded from the indexing and search processes. It's status changes to valid if it hits the market again. By reusing the product's information and image features, the indexing's performance is significantly improved.

**Search**. The search sub-system uses a distributed hierarchical architecture for scalability (Figure 1). It has three key components: Blender, Broker and Searcher, each of which has multiple instances for parallel processing and fault tolerance. The general workflow can be described as follows. Upon receiving a query from the user, a front end (i.e., load balancer) forwards the query to one of the blenders. The blender then sends the query to all brokers and each broker asks a subset of searchers to perform search in parallel. Each searcher is responsible for searching similar images from a partition of the entire image set. The searcher returns the top k most similar images to the requesting broker. The broker then combines the results from its subset of searchers and sends it back to the blender. The blender ranks the results and returns them to the user. The three level architecture offers scalability to large numbers of images, indexes and searches.

### 2.2 Full Indexing

The full indexing is performed periodically to ensure the data completeness. The process is shown in Figure 2.

All product update messages of a day are buffered in a message log. At the end of the day, each message in the log is processed in order. The images of new added products during the day are pulled from an image store and their high dimensional features are extracted. The feature extraction process first checks if the image's features have been extracted through a distributed key-value store. If it is a new image, the features are extracted and stored in the feature database. The feature database contains each image's high dimensional features and its corresponding product's attributes. The product attributes such as product ID, sales, prices and image URL are used to search and rank results. A bitmap file is used to indicate if an image URL is valid and hence determines if the image is indexable or not. The bitmap is updated according to the product information that is received from the message queue. For example, removing a product off the market makes its images invalid and vice versa. Only the valid images are used to create the full index. This optimization can significantly improve the indexing and search's performance.

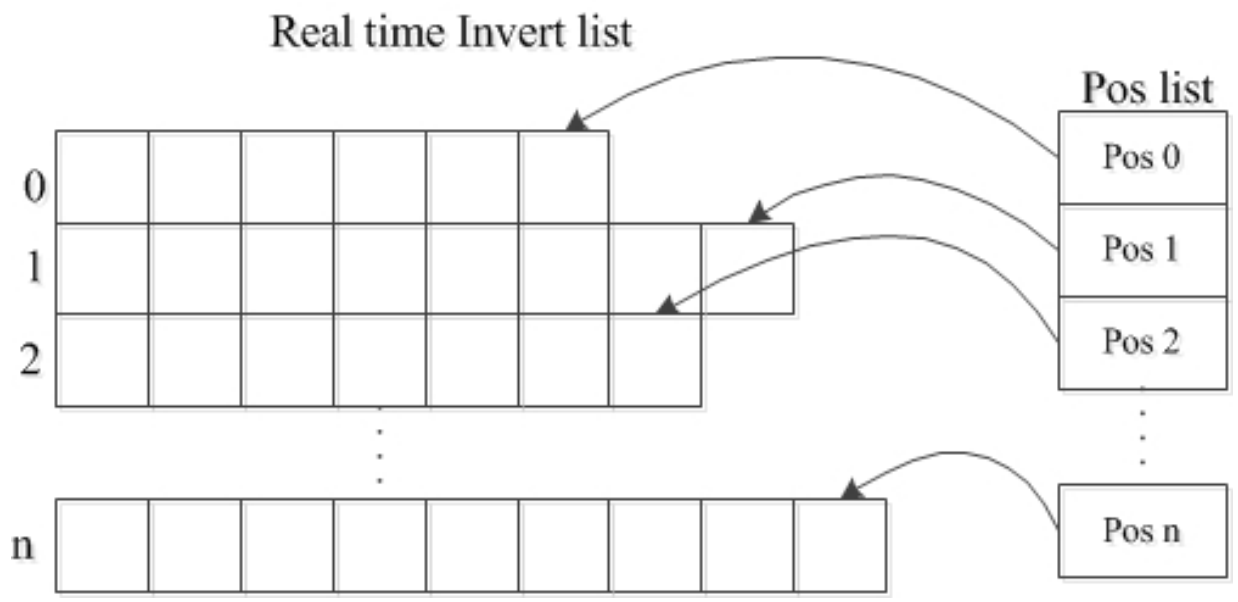

**Figure 5. Real Time Inverted Index Lists**

Building the full index for all images is performed every week. The process is shown in Figure 3. During the process, a forward index and an inverted index are created.

Each image is numbered sequentially and the product attributes of the image are stored in a forward index, which is a custom array and each element in the array contains the corresponding product's attribute information. The numeric attributes such as product ID, sales, price are stored in the fixed-length fields in the array. The variable length attributes like URL are stored in an additional buffer, and the offset of the attribute in the buffer is recorded in the array.

The inverted index is composed of N inverted lists. Each inverted list represents a class of images with similar high-dimensional features. The k-mean algorithm on a set of training data set (i.e., image features) is used to generate the classification. During the indexing process, the class that an image belongs to is calculated based on the similarity using the nearest neighbor algorithm and the image ID is appended to the corresponding inverted list.

### 2.3 Real Time Indexing

Figure 4 shows the real time indexing process. Messages about product or image updates are received from a message queue and processed instantly. In order to support real time update, there is an auxiliary array for storing the position of the last element in each inverted list (Figure 5). According to the message types, different operations are conducted as shown in Figure 6.

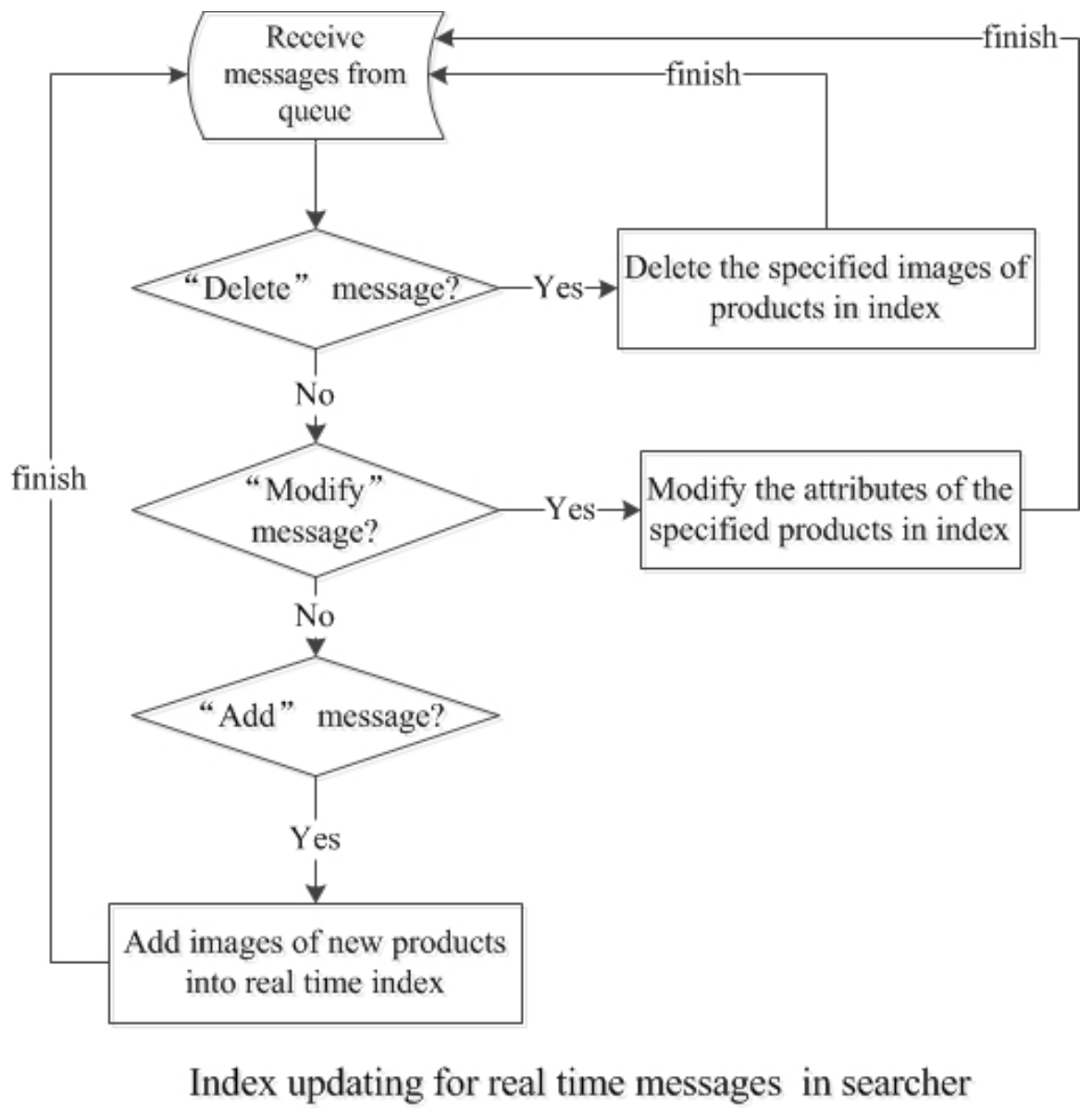

**Figure 6. Real Time Index Update Process**

**Update**. If the message is about a product's numeric attributes changes, the associated images' attributes in the forward index are updated (Figure 7). This operation is atomic and there is no conflict between search and update processes for maximum concurrency. For an attribute with varying length, the value is added at the end of the buffer and the offset value is updated in the forward index. If the availability of the product changes, for each image of the product, a flag in a bitmap indicating the image's validity is updated accordingly.

**Insertion**. If the message is about adding a product, we first check if the product already exists. If it is, we simply update its validity in the bitmap and reuse its images' features. If it is a new product, for each image of the product, a new index element plus the product's attributes are created in the forward index. The image URL is then inserted to the buffer and the offset is recorded in the forward index. To update the inverted index, the image's features are extracted and the inverted index list that the image belongs to is calculated based on its high-dimensional features. The image ID is then added to the end of the inverted list and the last element position of the inverted list is updated in the auxiliary array. The process is shown in Figure 8.

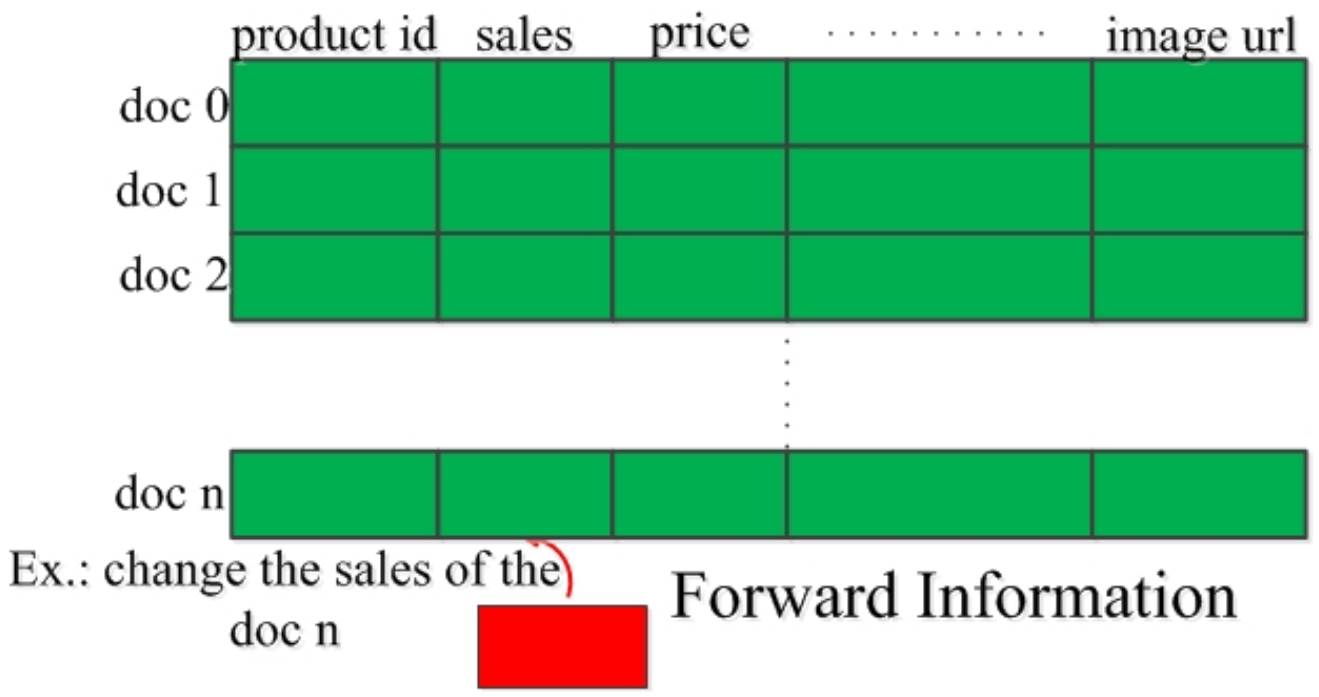

**Figure 7. Real Time Indexing: Update**

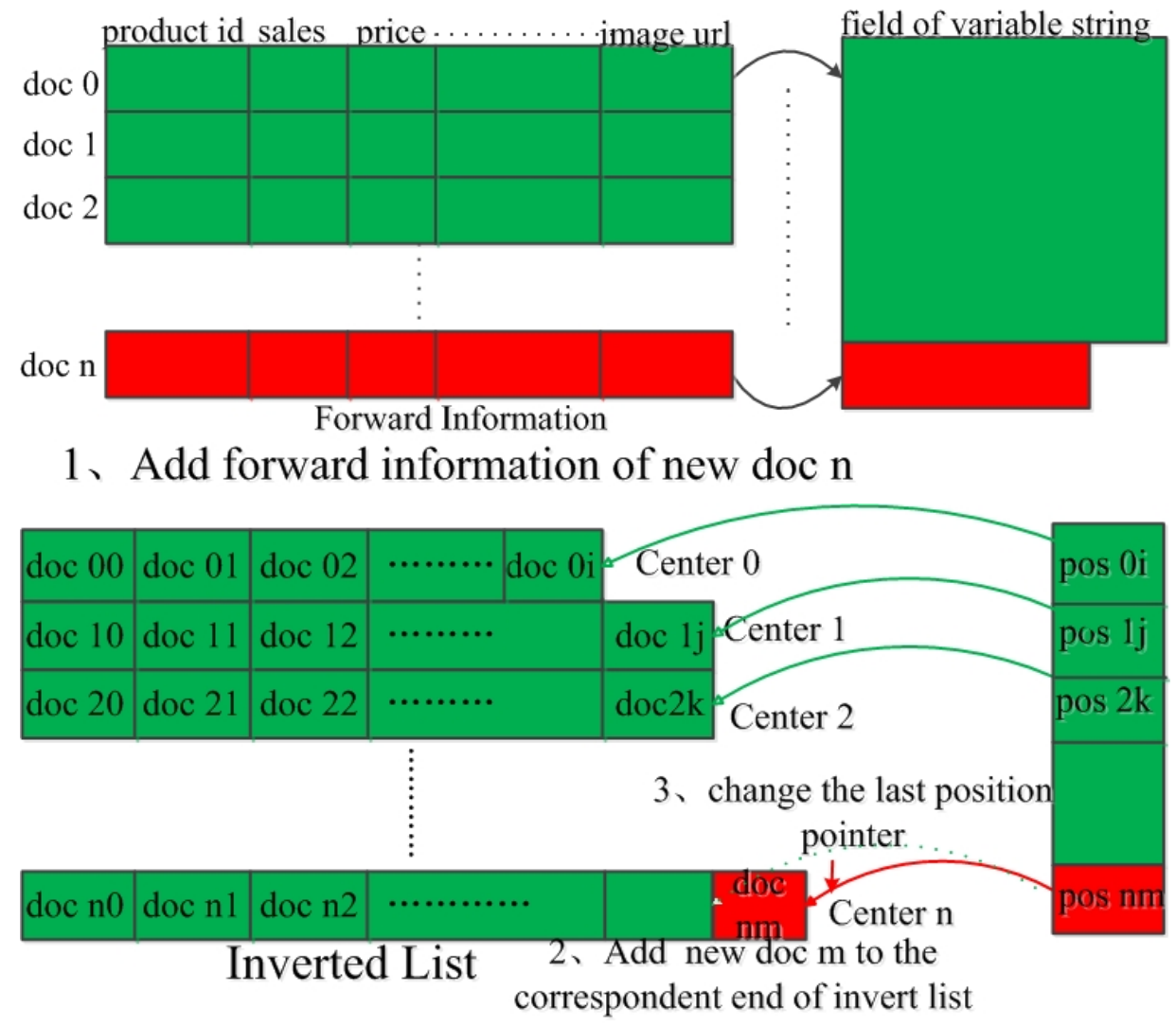

**Figure 8. Real Time Indexing: Insertion**

**Deletion**. Removing a product is as simple as changing the corresponding validity flag in the bitmap from 1 (valid) to 0 (invalid).

**Memory Management**. To accelerate the query, the memory of an inverted list is pre-allocated. When the current memory capacity is reached, a new inverted list of double size is created. The new image ID is added to the new created inverted list. The current inverted list continues to serve the requests until a background process finishes copying all the content of the current list to the new list. When the copy operation completes, the newly created inverted list becomes the current one and the old one is deleted. It ensures an lock-free and fast index update. Figure 9 shows the entire process.

### 2.4 Online Search

To search a picture, an item in the picture is detected and the product category of the item is identified. The high-dimensional features of the item's image is extracted. The most similar items (e.g., top k) are identified by traversing the inverted lists and calculating its Euclidean distance to each image in the inverted lists. Finally, the similar products are ranked according to their sales, praise, price and other attributes.

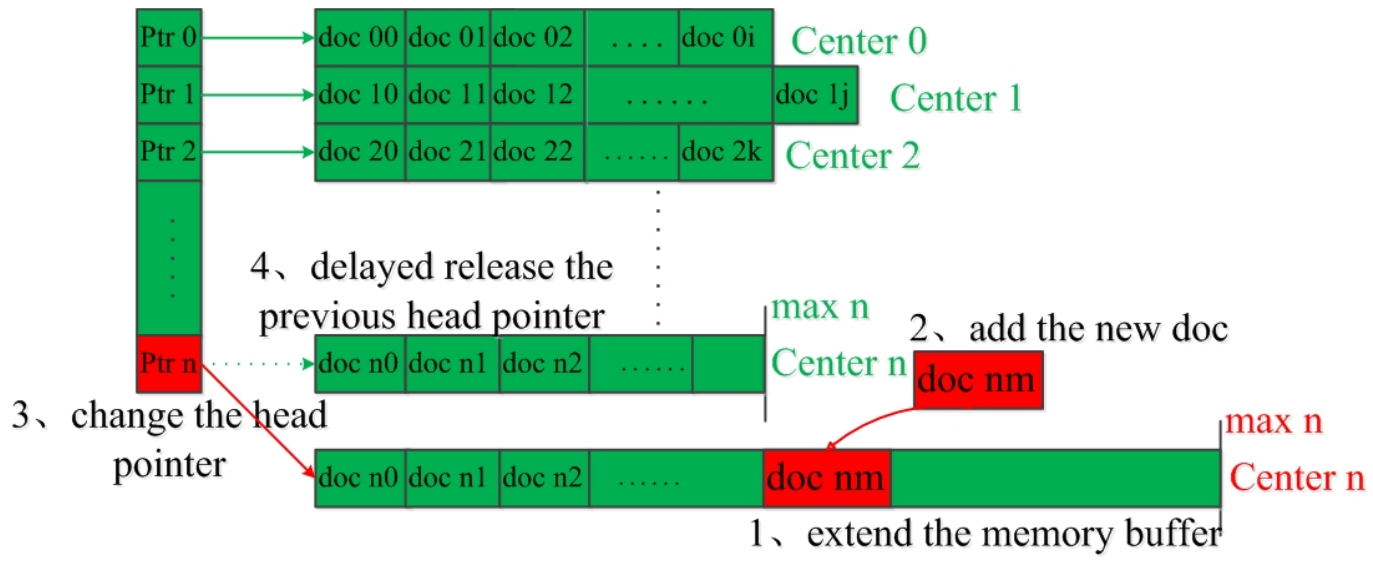

**Figure 9. Inverted List Expansion**

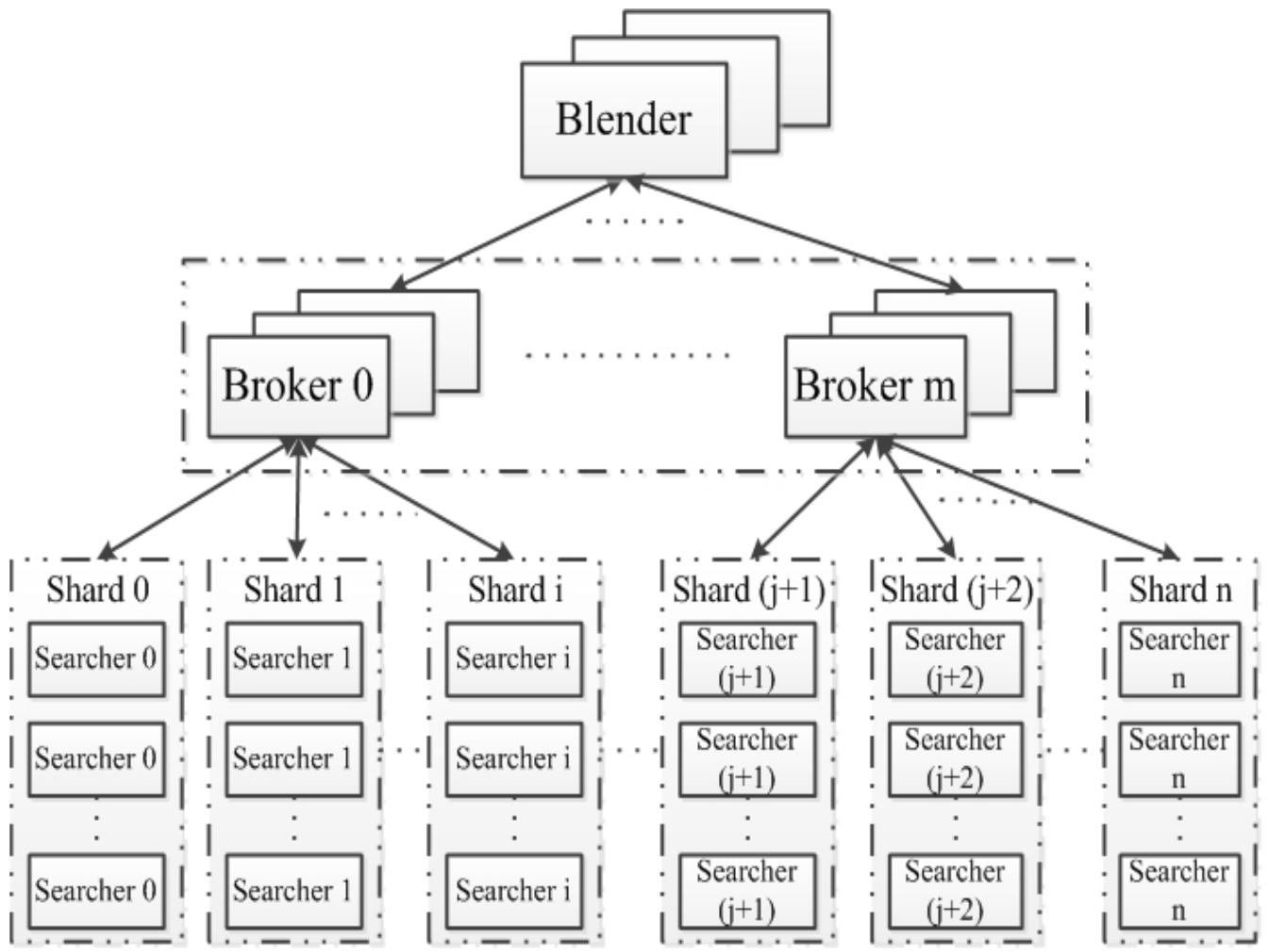

**Figure 10. Online Search Architecture**

Figure 10 depicts the detailed design of the online search sub-systems consisting of Blenders, Brokers and Searchers. The entire image index data is divided into multiple partitions by hashing the image's URL. Each partition can have multiple copies for availability. A partition is handed by a single searcher node. A broker connects to a subset of searchers. The 3-level structure ensures the system's scalability with a massive amount of data and a huge number of searcher nodes.

When a blender receives an image query request., it extracts the features and sends them to all the brokers. The blender also combines and ranks the results and returns to the user. A broker forwards the query to all the searchers it connects to and collects the partial search results from each searcher. Each broker has multiple identical instances for load balancing and fault tolerance.

There is a searcher for each index data partition. A searcher is responsible for searching and updating the corresponding index partition. Specifically, each searcher node identifies the cluster that is most similar to the queried image based on its features. It then scans the cluster's inverted list and calculates the similarity as each image in the inverted list. The top N most similar images are returned. A searcher is also responsible for processing messages from the message queue and performs real time indexing as described in Section 2.3.

## 3 Evaluation

### 3.1 Operation Data

The visual search system has been deployed on JD.com e-commerce site and used by JD's 300 million active users every day. Table 1 summarizes the operation data on August 4, 2018 collected from the production environment.

| Total | Attribute Update | Image Addition | Image Deletion |
|---|---|---|---|
| 977 million | 315 million | 521 million | 141 Million |

**Table 1. Number of Image Updates on 8/4/2018**

The system contains about 100 billion images and their feature indexes. The visual search system processed a total of 977 millions of image updates on that day. Among them, 315 millions were attribute updates, 521 millions were image additions, and 141 millions were image removals. Among the 521 millions of image additions, about 513 millions belong to the products which were removed from the market and put back again. These images' features were extracted before. We hence reused them to update the real index and significantly improved the response time. For the rest of the images, the feature extraction operations were performed.

Figure 11(a) shows the hourly rates of different types of real time index updates on August 4, 2018. The peak rate is about 80 millions/hour at 11am. The latency is shown in Figure 11(b). The average latency over 24 hour is 132ms. The average 99 percentile and 90 percentile latency are 816ms and 223ms perceptively.

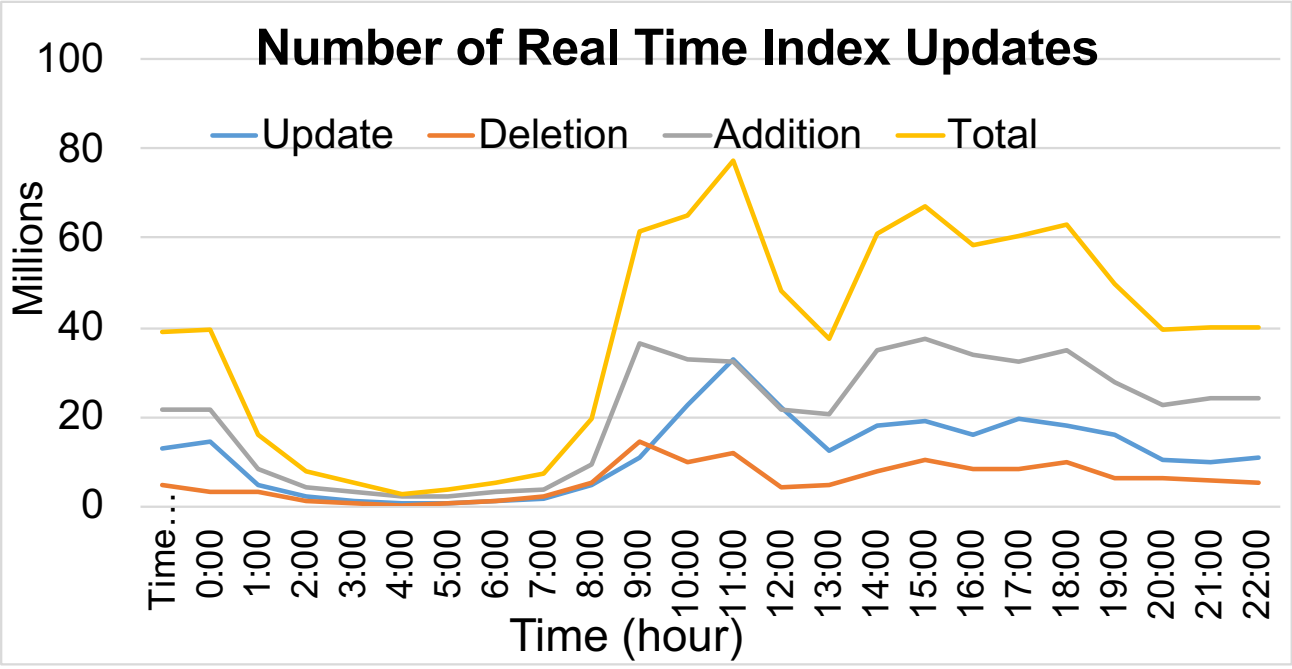

**(a) Hourly Rate of Real Time Indexing**

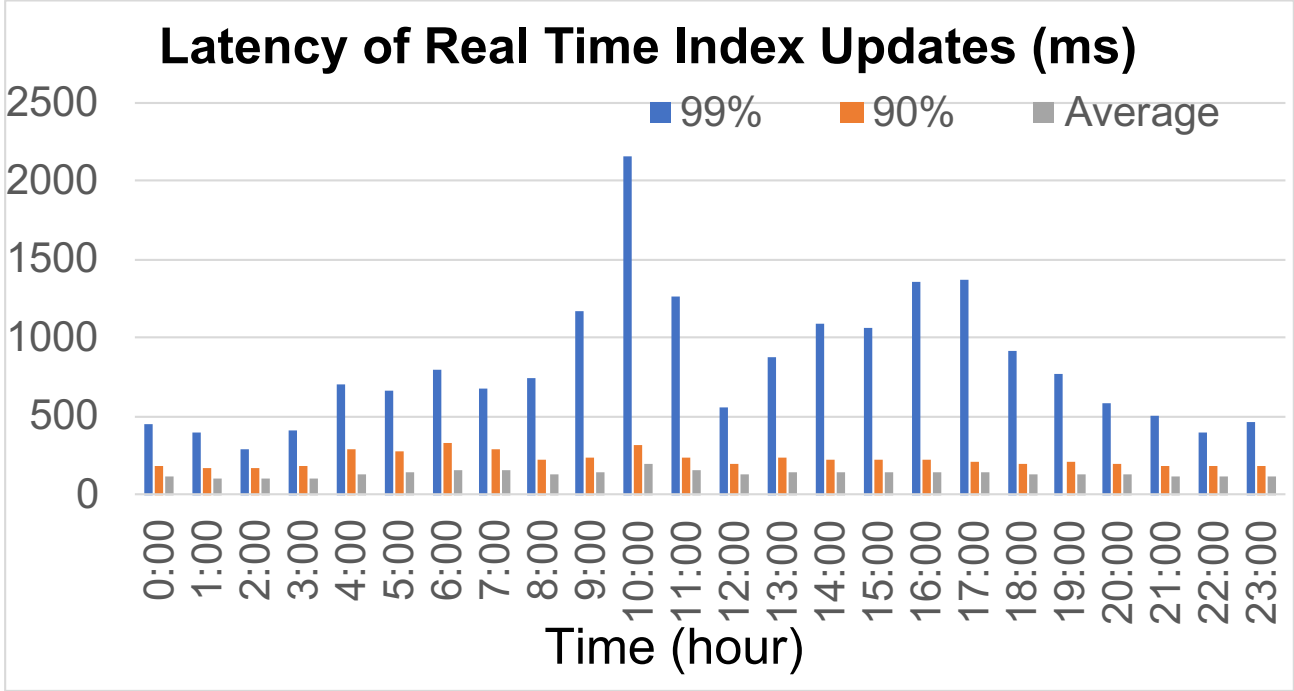

**(b) Performance of Real Time Indexing**

**Figure 11. Operation Data on August 4, 2018**

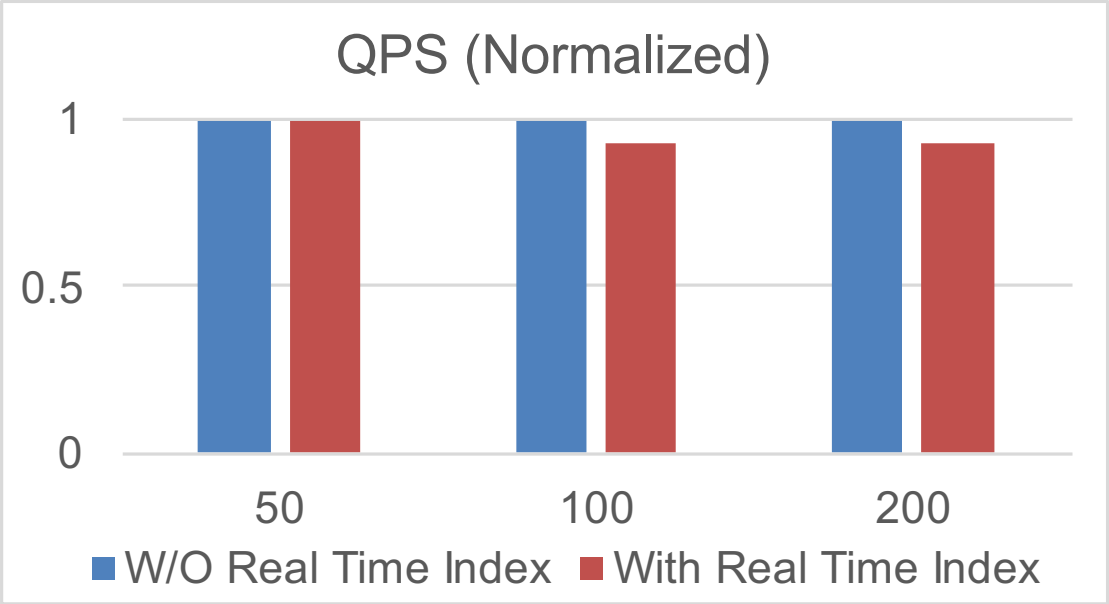

**(a) Throughput**

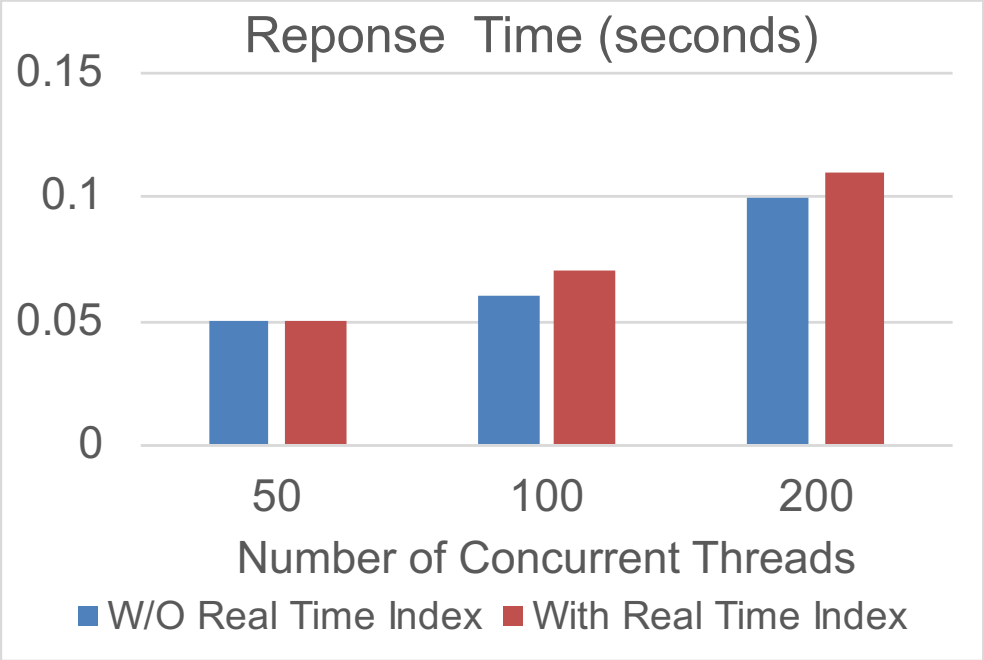

**(b) Response Time**

**Figure 12. Performance W/ and W/O Real Time**

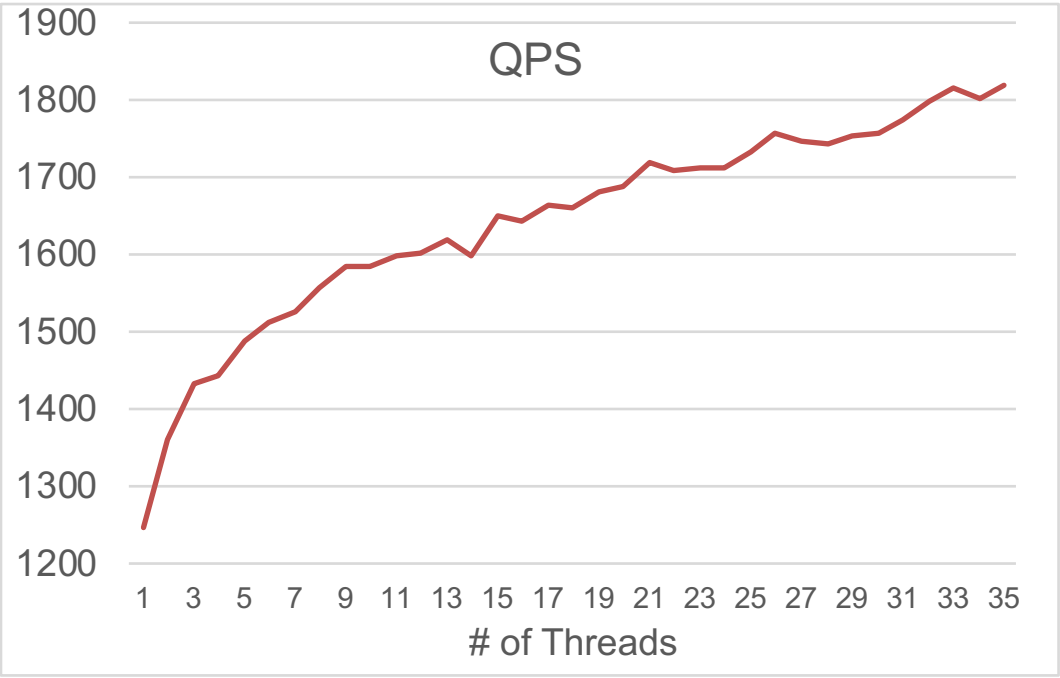

**(a) Throughput**

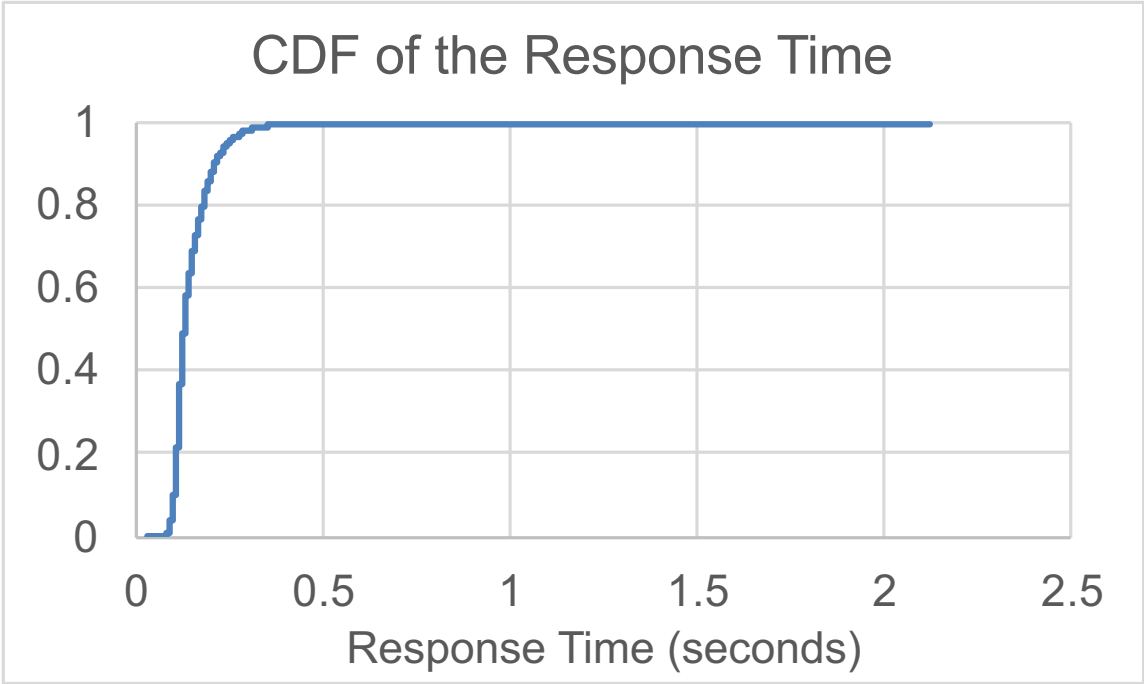

**(b) Query Response Time Distribution**

**Figure 13. Query Performance Scalability**

## 3.2 Performance Evaluation

We run performance experiments to evaluate the effectiveness of the design and implementation. The testbed consists of 1 server as the client machine for workload generation, 1 Nginx server as the front end, 6 servers as blenders and brokers, and 20 servers as searchers. Each server has 24 CPU cores and 256GB memory. A total of 100,000 images are used. The client machine emulates a different number of concurrent users by sending image query requests to the visual search system.

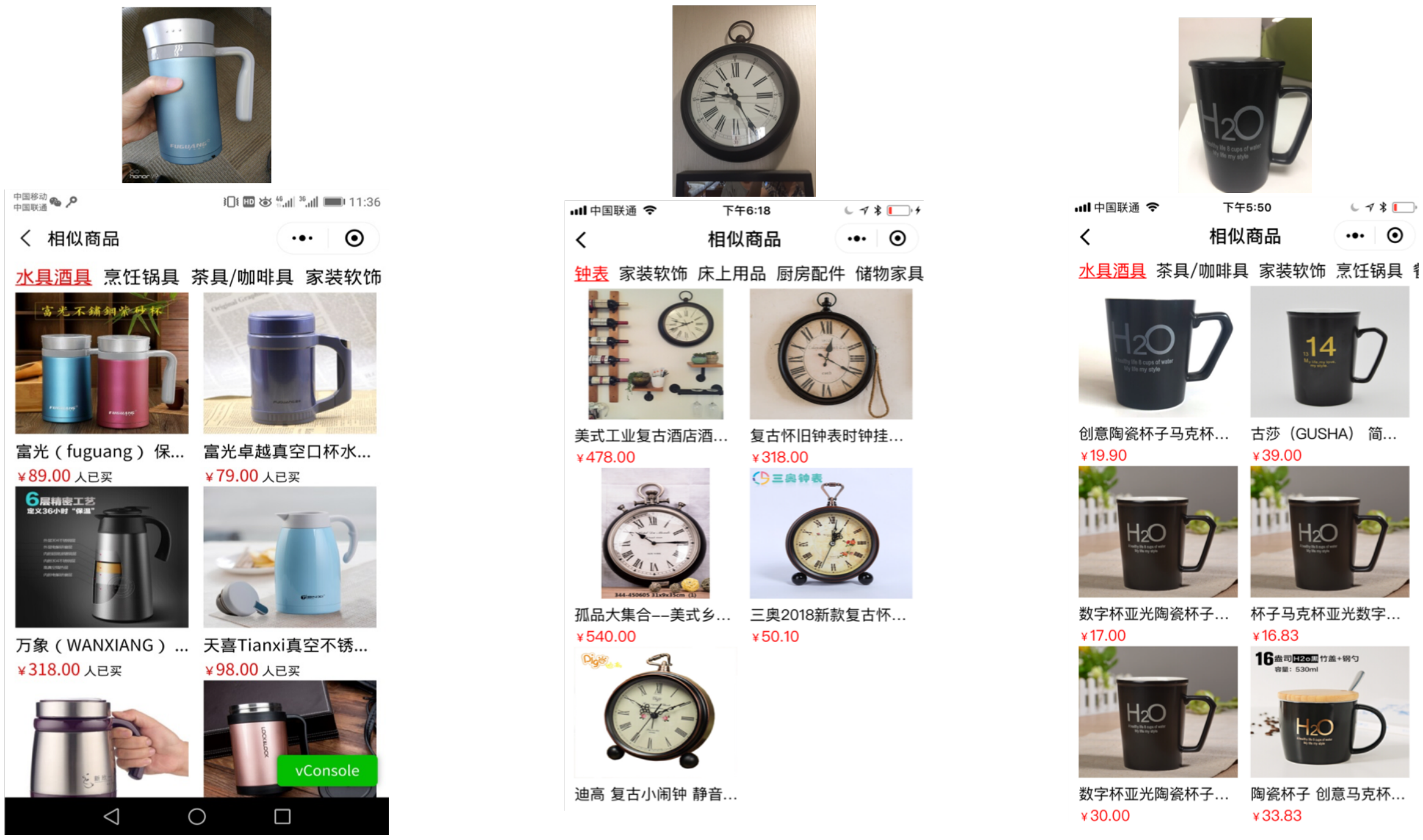

**Figure 14. Real Search Examples on a Mobile Application**

A key requirement of e-commerce visual search is data freshness, which is achieved via real time update and index in our system. Figure 12 compares the performance using real time index and without using real time index. As shown in the Figure 12(a), the throughput with real time index is comparable with that without real time index and the performance overhead introduced by real time index is less than 10%. The query response times with and without real time index are similar too shown in Figure 12(b). The average latency is less than 100ms, which is good enough for real time image search in practice. These results clearly demonstrate the effectiveness of the proposed real time indexing.

Figure 13 shows the scalability of the system. Figure 13(a) shows that the setup can support up to 1800 queries per second or 155-million search requests per day, which is more than enough to serve the visual search demand of most e-commerce sites. Figure 13(b) shows the query response time distribution at the maximum throughput. The maximum response time is 2.1 seconds and the 99 percentile response time is 0.3 seconds. The results demonstrate the system's ability to support real time visual search with the real production system.

### 3.3 Search Examples

The visual search system has been implemented and deployed as part of JD.com shopping application. Figure 14 shows three real search examples on a mobile phone application, which returns the top 6 similar products for three different searches.

## 4 Related Work

Image content-based search has been studied extensively in the last 10 years. A considerable number of algorithms have been developed [5,8,14-17,23-30]. Similar algorithms are used in our system, but the focus of this paper is on the system design and implementation. Many image web search systems aimed to address the scalability challenges [9-13], but real time update and data freshness are often not their focus. The visual search systems at Pinterest, Alibaba and eBay [1-4] are similar to ours in terms of performing product image search. However, they mainly looked at the algorithm design and optimization, and didn't address the system issues such as real time indexing and update as our work. Efficient indexing was studied in [21,22], but neither addressed the real time issues.

## 5 Conclusion and Future Work

In this papers, we demonstrate how to design and implement a real time visual search system on JD's Internet-scale e-commerce site through distributed architecture and efficient indexing. We plan on integrating advanced search and ranking algorithms into our visual search system in the future work.